\documentclass{sigchi-ext}
\usepackage[T1]{fontenc}
\usepackage{textcomp}
\usepackage[scaled=.92]{helvet} 
\usepackage{graphicx} 
\usepackage{balance}  
\usepackage{booktabs} 
\usepackage{ccicons}  
\usepackage{ragged2e} 
\usepackage{todonotes}

\usepackage{marginnote} 

\copyrightinfo{
{\emph{CHI'20} Workshop on Human-Centered Approaches to Fair and Responsible AI
}, 26th of April, 2020, Honolulu, HI, USA. Copyright is held by the authors.}

\def\plaintitle{Exploring User Opinions of Fairness in Recommender Systems} 
\def\emptyauthor{}
\def\plainkeywords{fairness; multistakeholder recommendation algorithms; consumers; providers; trust; bias;  transparency; explanation; user study;}

\title{Exploring User Opinions of Fairness in Recommender Systems}

\numberofauthors{4}
    
\author{%
  \alignauthor{%
    \textbf{Jessie Smith}\\
    \affaddr{Department of Information Science, University of Colorado} \\
    \affaddr{Boudler, CO}\\
    \email{jessie.smith-1@colorado.edu} 
     }
    \alignauthor{%
    \textbf{Casey Fiesler}\\
    \affaddr{Department of Information Science, University of Colorado}\\
    \affaddr{Boudler, CO}\\
    \email{casey.fiesler@colorado.edu}
      }\vfil
    \alignauthor{%
    \textbf{Nasim Sonboli}\\
    \affaddr{Department of Information Science, University of Colorado}\\
    \affaddr{Boudler, CO}\\
    \email{nasim.sonboli@colorado.edu} } 
    \alignauthor{%
    \textbf{Robin Burke}\\
    \affaddr{Department of Information Science, University of Colorado}\\
    \affaddr{Boudler, CO}\\
    \email{robin.burke@colorado.edu}  } 
    }

\definecolor{linkColor}{RGB}{6,125,233}
\hypersetup{%
  pdftitle={\plaintitle},
  pdfauthor={\emptyauthor},
  pdfkeywords={\plainkeywords},
  bookmarksnumbered,
  pdfstartview={FitH},
  colorlinks,
  citecolor=black,
  filecolor=black,
  linkcolor=black,
  urlcolor=linkColor,
  breaklinks=true,
}

\begin{document}


\maketitle

\RaggedRight{} 

\begin{abstract}
  Algorithmic fairness for artificial intelligence has become increasingly relevant as these systems become more pervasive in society. One realm of AI, recommender systems, presents unique challenges for fairness due to trade offs between optimizing accuracy for users and fairness to providers.
  But what \emph{is} fair in the context of recommendation--particularly when there are multiple stakeholders? In an initial exploration of this problem, we ask users what their ideas of fair treatment in recommendation might be, and why. We analyze what might cause discrepancies or changes between user's opinions towards fairness to eventually help inform the design of \emph{fairer} and more transparent recommendation algorithms.
\end{abstract}


\printccsdesc

\section{Introduction}

Historically the term \emph{fairness} has been difficult to define. People's opinions of what constitutes fair or unfair treatment differ between cultures and throughout time \cite{reichart:ethics}. As algorithms become more deeply embedded into social contexts like education, healthcare, policy, and the internet, the issue of defining fair treatment is also increasingly interdisciplinary. In the discipline of computer science - and more specifically  machine learning - researchers have begun to tackle these problems. 

Previous work has made an effort to turn philosophical definitions of fairness into metrics that machine learning models can optimize for \cite{gajane2017formalizing,barocas-hardt-narayanan}; however, these definitions may not be sufficient for diverse users and groups \cite{sociotechnical}. In this project we focus on one specific realm of machine learning, recommender systems with multiple stakeholders, as we explore what it means for these algorithms to be fair or unfair - from the viewpoint of those who consume the recommendation. \emph{Multistakeholder} recommender systems are unique to issues of fairness, because of an inherent tradeoff for fairness between multiple sets of users~\cite{abdollahpourimulti2020}. For example, two types of users for a multistakeholder recommendation algorithm could be (1) those who \emph{provide} items to be recommended (providers), and (2) those who consume the recommended items (consumers).

\marginpar{%
  \vspace{-218pt} \fbox{%
    \begin{minipage}{0.925\marginparwidth}
      \emph{\textbf{P15}: "If Zillow is systematically recommending that certain subgroups of the population rent houses in dangerous neighborhoods... an obvious disadvantage for that group of people. If indeed.com is recommending jobs for me that are not reflective of my full potential as a candidate and it is because of some bias in their recommendation system, that's not fair to whatever person is receiving these job alerts... I think that those are all things that have larger consequences than what movie you watched tonight or what song you listened to on your iPhone."}
    \end{minipage}}\label{sec:sidebar1} }
    
\marginpar{%
  \vspace{-1pt} \fbox{%
    \begin{minipage}{0.925\marginparwidth}
      \emph{\textbf{P11-1}: "That changes my views a little bit based on the goals of the company and what they are trying to get out of these recommendation algorithms, because I think nonprofits are more about just trying to connect you to the things that would interest you more... as opposed to companies trying to generate more money based on your preferences and your views."}
    \end{minipage}}\label{sec:sidebar2} }
    

In the case of a company like Kiva, a micro-lending platform, the providers would be borrowers who are seeking funding for their loans, and the consumers would be those who are looking to lend money to others. If Kiva began to provide loan recommendations for consumers, there would need to be a decision for how \emph{fair} the recommender system should be for the providers. In this example, provider-fairness is a type of recommendation diversity. As recommendations represent a more diverse set of providers (e.g., both over-funded and under-funded borrowers on Kiva), they tend to become less personalized (less accurate) for the consumer \cite{liu2019personalized}. Thus, recommendation algorithms with multiple stakeholders will need to draw a line between how diverse versus how personalized the system should be. It is apparent that consumers and providers will have different opinions about where this line should be drawn. In this work, we ask the consumers for their opinion.

\section{What do the Consumers Think is Fair?}

For this exploratory study, we conducted interviews with 30 participants (majority college students) in which we asked them to reflect on fairness issues in the context of recommender systems, using both systems they are familiar with (e.g., Netflix, Amazon) and Kiva as examples. We analyzed our data using thematic analysis \cite{thematicanalysis}, and arrived at a number of overarching themes, a subset of which we discuss here.

\subsection{\textbf{I. Consequences of the System}}

First, participants tended to want more provider-fairness (diversity) and less accuracy (personalization) when they recognized that recommendations could have a noticeable, harmful effect on certain stakeholders in the system. Though most participants preferred less personalization as provider risk became higher, none indicated wanting to completely omit personalization. In the context of recommendations, this viewpoint makes sense on platforms like Netflix or Spotify, where the user expects some level of personalization to derive utility from the platform. However, most participants still expected accuracy for recommendations that do not always require personalization, such as popular/trending items, which struggle with issues of provider-fairness as well.

Many participants expressed that certain kinds of recommendations (such as housing, job recommendations, healthcare, or finances) should include fairness as a central goal, due to  potential for harmful consequences in an unfair system. For example, P15 contrasted Netflix and Spotify with Zillow and Indeed.com, noting that for the former it doesn't really \emph{"matter,"} but it would, e.g., for housing or employment (see sidebar).

\subsection{\textbf{II. Nonprofits Versus For-Profit Companies}}

Another important influence was the kind of organization that was providing  recommendations. Specifically, participants tended to trust nonprofit fairness goals (e.g., Kiva's) more than for-profit companies, which led them to indicate a preference for less personalization and more diversity on nonprofit platforms. For example, multiple participants described differences between for-profit companies and non-profits, in terms of both motives and consequences (see sidebar for examples from P11-1 and P8).

    
\marginpar{%
  \vspace{-100pt} \fbox{%
    \begin{minipage}{0.925\marginparwidth}
      \emph{\textbf{P8}: "Kiva has really big consequences if someone doesn't get that funding, you know, versus Amazon doesn't have as big of a consequence cause Amazon's more detached thoroughly. Like if you buy it, that's great, they get more money. But if you don't buy it... they're still getting money. Versus Kiva, if you buy it, that's great, somebody gets clean water, if you don't buy it, somebody is not getting clean water."}
    \end{minipage}}\label{sec:sidebar3} }
    
\marginpar{%
  \vspace{1pt} \fbox{%
    \begin{minipage}{0.925\marginparwidth}
      \emph{\textbf{P22}: "I feel like right now, the way things are, it's kind of capitalistic and promotes the wealthy getting wealthier. And people who are struggling to start off a business, and maybe failing - with a fairness goal it would be better."}
    \end{minipage}}\label{sec:sidebar4} }
    
\marginpar{%
  \vspace{1pt} \fbox{%
    \begin{minipage}{0.925\marginparwidth}
      \emph{\textbf{P11-2}: "I think a fair algorithm, if such a thing is even possible, would be something that really, I guess is non-biased. But of course, everything's biased in some way."}
    \end{minipage}}\label{sec:sidebar5} }

\subsection{\textbf{III. Bias in Prioritization for Multistakeholder Systems}}

Just as philosophers have debated how values and definitions of fair treatment differ between communities \cite{flanagan-morals}, researchers too have discovered that people who come from different backgrounds have different opinions towards algorithmic fairness \cite{demographics-fairness}. In this study, we noticed that the varying degrees in which participants were willing to give up personalization for provider-fairness were heavily influenced by their personal biases or predispositions. For example, P22 - who had experience as a provider selling items on Etsy - empathized more with other sellers and thus wanted more provider-fairness over personalization (see sidebar). In one case, P11-2 alluded that stakeholder bias might be the ultimate obstacle in creating a fully fair recommendation algorithm (see sidebar). It is of course inevitable that designers of fairness-aware recommender systems will tend to include their own personal biases into decisions concerning their platform's fairness goals.

Multiple participants expressed discomfort when choosing whether to prioritize the consumer or the provider for recommendations on the Kiva platform, where the risk was much higher. For example, on Kiva's microlending platform choosing more personalization for recommendations could increase consumer biases (e.g., implicit bias in lending selection), but choosing greater diversity could increase the potential for algorithmic bias (e.g., popularity bias leaving certain borrowers consistently underfunded). P12-1 explained their desire to utilize platform design to understand and combat their own personal biases (see sidebar).

\subsection{\textbf{IV. Transparency / Explanations}}

Explanation also plays a role in opinions towards fair treatment in recommendations. Recent work has highlighted that people's perceptions of algorithmic justice are altered when the algorithm's decisions are explained in different ways \cite{justice-algorithms}, and in our study participants indicated that while they would like to have some transparency through explanation when recommendations are altered for provider-fairness, they did not want the explanations to be meant to convince the consumer to change their mind, as described by P12-2 (see sidebar).

\section{Conclusion and Future Work}
Overall, this work is a starting point to build a better understanding of where to draw the line between recommendation personalization and provider-fairness in multistakeholder recommender systems. While it is important to keep in mind the preferences of the consumer, as this study has done, future work could dive into the preferences of providers (such as Spotify musicians or Amazon sellers), as well as the preferences of the designers of these systems. In multistakeholder environments, it is very important to appropriately balance the interests of every member of the system in order to build trust, maintain accuracy, and promote equality. The indication that many consumers have different preferences for recommendation fairness is somewhat alarming, but also important evidence that this work is necessary to ensure greater fairness in the future, and to understand the necessary tradeoffs. A greater understanding of what stakeholder's preferences are will allow for more transparency of these tradeoffs in the future, to ensure that every stakeholder's interests in recommender systems are taken into consideration.

\marginpar{%
\vspace{-30pt} \fbox{
    \begin{minipage}{0.925\marginparwidth}
      \emph{\textbf{P12-1}: "I think that it would be more successful if recommendations were defaulted to fairness with the choice to undo it. I think it can really help people understand their own bias as well. And being more aware, like, oh, I do want to spend my money on this instead of this, like I originally thought...So I think communicating to people why it's happening would be really helpful."}
    \end{minipage}}\label{sec:sidebar6} }
    
\marginpar{%
  \vspace{-1pt} \fbox{%
    \begin{minipage}{0.945\marginparwidth}
      \emph{\textbf{P12-2}: "I think [explanations] can definitely push people to make decisions they weren't expecting to make. Sometimes that can be good if that's creating equity for the users and the sellers on the website. But if there is not fairness built into an algorithm and it's pushing you to make these decisions, it can almost brainwash you into thinking a certain way without knowing it. That's the scary part."}
    \end{minipage}}\label{sec:sidebar7} }

\balance{}

\bibliographystyle{SIGCHI-Reference-Format}
\bibliography{sample}

\end{document}